# The Effect of short-range order on the viscosity and crystallization of Al–Mg melts

Elizaveta A. Batalova[1], Larisa V. Kamaeva[1,2] and Nikolay M. Chtchelkatchev[2]

[1] Udmurt Federal Research Center, Ural Branch of Russian Academy of Sciences, 426068 Izhevsk, Russia
[2] Vereshchagin Institute for High Pressure Physics, Russian Academy of Sciences, 108840 Troitsk, Moscow, Russia

E-mail: lara_kam@mail.ru

**Abstract**

In this work, using the methods of viscosimetry and thermal analysis, the concentration changes in the values of the supercooling viscosity of Al–Mg melts with Mg content from 2.5 to 95 at.% are studied. It is shown that the temperature dependences of viscosity are well described by an exponential dependence. The concentration dependence of viscosity is not monotonous and reflects a change in the chemical short-range order in the liquid phase. The concentration dependence of supercooling of Al-Mg melts is determined by the type of solid phase formed during solidification, and also reflects the most significant changes in the chemical short-range order in the liquid phase at 20 and 80 at.% Mg. Al-Mg alloys in the concentration ranges: 0-10, 40-50 and 90-100 at.% Mg are prone to non-equilibrium crystallization, the formation of quasi-eutectics and solidification without intermediate intermetallic phases.

## 1. Introduction

Al-based metal systems have the complex structure liquid state [1–3]. However, these alloys are actively used as structural materials. In particular, the Al–Mg alloys are widely used in aircraft, shipbuilding, and mechanical engineering. This is due to the fact that adding a small amount of Mg to Al leads to a simultaneous increase in the strength and ductility of the obtained alloys [4]. Among the advantages of Al–Mg alloys are also formability, weldability, and high corrosion resistance associated with the formation of a poorly soluble oxide film on the surface. On the other hand, Al-Mg alloys are characterized by susceptibility to intergranular corrosion, which is due to the appearance of β-phase ($Al_3Mg_2$) crystals in the structure of the alloys; therefore, the conditions for its formation are intensively studied [5]. For a wider practical application of Al-Mg alloys in the range of Mg-rich concentrations, it is necessary to understand the mechanisms of formation of complex compounds: $\gamma$-$Mg_{17}Al_{12}$, $\zeta$-$Al_{52}Mg_{48}$ [6] and $\varepsilon$-$Al_{30}Mg_{23}$ [7]. An analysis of the electronic structure and vibrational entropy of intermetallic compounds showed that the description of the formation of such compounds requires an analysis of the short-range order in the arrangement of atoms [9-11]. The short-range order in the liquid and supercooled Al–Mg melts with Mg content from 0 to 100 at.% using ab initio molecular dynamics (AIMD) was carried out in [3–12]. After analyzing the Warren-Cowley parameters, the authors showed that due to the formation of extended bonds between aluminum atoms and the fragmentation of magnesium atomic groups in the concentration ranges with a magnesium content of 20-25 and 70-80 at.%, changes in the chemical local ordering are observed. Our studies [13–16] for Al-based ternary systems showed that concentration changes in the chemical short-range order in melts affect not only the concentration behavior of properties, but also the crystallization process, which confirms the structural heredity between the liquid and solid states. Therefore, taking into account the identified features of the chemical short-range order in [3, 12], in this work we have studied the kinematic viscosity and crystallization processes of melts of the Al–Mg system with magnesium content from 2.5 to 95 at.%.

## 2. Materials and methods

The kinematic viscosity (ν) of liquid Al–Mg alloys was measured on an automated setup [17] using the method of damped torsional vibrations of a crucible with a melt in the Shvidkovsky variant [18]. Samples for research weighing 8-12 g were obtained by melting electrolytic aluminum and high-purity magnesium in the furnace of the installation [17] in a protective atmosphere of purified helium at a temperature of 800 $^0$C for 30 min. After remelting, viscosity polytremes of the studied melts were obtained in the stepwise cooling mode after isothermal exposures at each temperature for 5 minutes. Upon completion of the crystallization of the samples and their cooling to room temperature, the viscosity was measured in the mode of reheating and subsequent cooling in the same temperature ranges. The crucible and the cylindrical cover, which is the second end surface [19], were made of $Al_2O_3$. Two end surfaces of contact between the melt and the crucible are necessary to prevent the uncontrolled influence of the oxide film. All measurements were carried out in a protective atmosphere of purified helium.

Differential thermal analysis (DTA) was implemented on an automated high-temperature thermal analyzer (VTA-983). The method of operation on this setup is described in [20]. VTA-983 operates in the temperature range from 100 to 1700 $^0$C in an inert atmosphere of purified He under a slight excess pressure after preliminary evacuation to $10^{-2}$ Pa. The recorded parameter is the differential temperature, defined as the difference between the temperatures of the test sample and the reference. Tungsten is used as a standard. Samples for research weighing 0.4 g were obtained by alloying electrolytic aluminum and high-purity magnesium in a VTA-983 furnace at a temperature of 800 $^0$C for 15 min., then the samples were slowly cooled at a rate of 20 $^0$C/min. In order to obtain an accurate composition, when calculating the weights, the possible losses of magnesium during its high-temperature remelting and during the experiments were taken into account.

The graphs of differential thermal analysis (thermograms) in this work were obtained according to the following scheme: after remelting, the samples were heated at a rate of 20 $^0$C/min to a temperature of 800 $^0$C, kept at this temperature for 15 min., and cooled at a rate of 100 $^0$C/min. In this work, a series of experiments on cycling with different cooling rates and thermal cycling were carried out.

Two series of experiments are needed to analyze the influence of the melt temperature on the magnitude of supercooling ($\Delta T$).

Thermal cycling was carried out as follows: heating the sample to a predetermined temperature, holding for 10 minutes, cooling at a rate of 100 $^0$C/min. In the next cycle, the overheating temperature of the melt increases by 25-50 $^0$C, and so - up to a maximum temperature of 800 $^0$C. As a result, thermograms of heating and cooling from different temperatures were obtained, which were used to determine the equilibrium liquidus temperature (during heating) and the liquidus temperature during crystallization under cooling conditions from different temperatures. The corresponding subcooling was calculated as the difference between the above liquidus temperatures. As a result, dependences of supercooling, under which crystallization of the melt begins, on the initial temperature of the melt were obtained.

Серия измерений в экспериментах по циклированию проводилась с различными скоростями охлаждения (20, 50 и 100 $^0$C/мин.) от одинаковой температуры расплава. На концентрационных зависимостях переохлаждения изображены все полученные для каждого исследуемого сплава данные при различных условиях охлаждения. Это необходимо для получения объективных значений.

The structural characteristics of the studied melts were determined using first-principles molecular dynamics modeling (AIMD). The calculations were carried out using the Vienna ab initio simulation program (VASP) [21]. A detailed calculation procedure is presented in [13, 15].

## 3. Results and discussion

Typical temperature dependences of the viscosity of the studied Al–Mg melts (from 2.5 to 95 at.% Mg) in the coordinates $\nu(T)$ and $\ln\nu(1/T)$ are shown in Fig. 1. In the studied temperature range, significant anomalies in the temperature dependences of the viscosity of Al melts –Mg was not detected, which indicates that thermal treatment of these melts does not result in a sharp change in the atomic short-range order. Viscosity polytherms in heating and cooling modes are monotonic dependencies and are well described by the Arrhenius relation:

$$\nu = A_\nu \exp\left\{\frac{E_\nu}{RT}\right\}, \qquad (1)$$

where $A_\nu$ is a constant, $E_\nu$ is the activation energy of the viscous flow.

At temperatures near the melting point, polytherms deviate from the exponential dependence, which is due to the methodological features of measuring viscosity in the presence of second end surface [19]. The increased values of viscosity at low temperatures are associated with the penetration of the melt into the gap between the crucible and the lid as a result of the good wettability of the crucible material (Al$_2$O$_3$) by magnesium melts.

Table 1 shows the parameters of the approximating equation:

$$\ln(\nu) = a\frac{1}{T} - b \qquad (2)$$

**Table 1.** Parameters of the approximating equation (2) of the temperature dependences of the viscosity of Al-Mg melts.

| Composition, at. % | a | b |
|---|---|---|
| Al$_{97.5}$Mg$_{2.5}$ | 1309 | 15.99 |
| Al$_{95}$Mg$_5$ | 1365 | 15.90 |
| Al$_{92.5}$Mg$_{7.5}$ | 1559 | 16.06 |
| Al$_{90}$Mg$_{10}$ | 1348 | 15.79 |
| Al$_{87.5}$Mg$_{12.5}$ | 1547 | 15.97 |
| Al$_{85}$Mg$_{15}$ | 1577 | 15.96 |
| Al$_{80}$Mg$_{20}$ | 1708 | 15.94 |
| Al$_{75}$Mg$_{25}$ | 1561 | 16.11 |
| Al$_{70}$Mg$_{30}$ | 1902 | 16.16 |
| Al$_{60}$Mg$_{40}$ | 2115 | 16.31 |
| Al$_{55}$Mg$_{45}$ | 2076 | 16.21 |
| Al$_{50}$Mg$_{50}$ | 2180 | 16.31 |
| Al$_{40}$Mg$_{60}$ | 2134 | 16.22 |
| Al$_{32}$Mg$_{68}$ | 1953 | 16.03 |
| Al$_{30.5}$Mg$_{69.5}$ | 1820 | 15.90 |
| Al$_{28.4}$Mg$_{71.6}$ | 1919 | 16.06 |
| Al$_{25}$Mg$_{75}$ | 1526 | 15.66 |
| Al$_{20}$Mg$_{80}$ | 1216 | 15.31 |
| Al$_{10}$Mg$_{90}$ | 1804 | 15.86 |
| Al$_5$Mg$_{95}$ | 1751 | 15.87 |

Figure 2 shows the concentration dependences of the viscosity of the studied melts for different temperatures, built on the basis of the previously obtained polytherms. It can be seen from the figure that with an increase in the magnesium concentration in Al–Mg melts, their viscosity increases. Moreover, the increase occurs nonmonotonically: there are 5 intervals with different character of dependence $\nu$ (T). Viscosity increases sharply at a magnesium content of 2.5 to 5 at.%, when its concentration reaches 7.5 at.%, the same sharp decrease occurs, as a result, a pronounced maximum appears on the concentration dependence at 5 at.% Mg. With an increase in the magnesium concentration in the melt to 20 at.%, an increase in viscosity occurs, with a magnesium content of 22.5 at.%, a pronounced minimum is observed, and then its monotonous increase occurs until the magnesium concentration reaches 72 at.%. Then there is a minimum at 75 at.% Mg, which corresponds to the eutectic concentration, after which the viscosity increases to 80-85 at.% Mg, and then begins to decrease again.

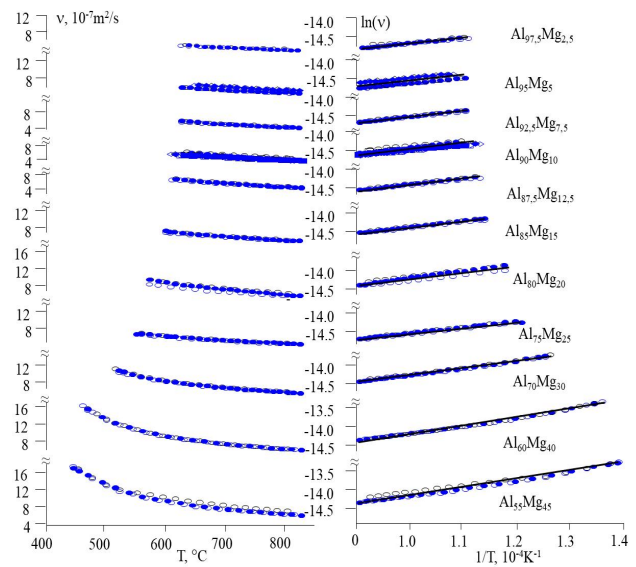

**Fig.1** Temperature dependences of the viscosity of Al-Mg melts in direct and semi-logarithmic coordinates when measured in the isothermal holding mode for 5 min. With stepwise cooling (light circles) and subsequent heating (blue circles))

The concentration features of the viscosity of Al–Mg melts that we found are in good agreement with the data obtained in [12] on the concentration change in the chemical short-range order in these melts. The authors of the above work used the Warren-Cowley coefficients, which show how much the real chemical nearest environment of the selected atom differs from the random one at a certain concentration of components. In order to study in more detail the processes occurring in the regions with magnesium content of 20 and 80 at.% (the largest changes in viscosity, which manifest themselves in its concentration dependence), we also analyzed the chemical local ordering by the method of ab initio molecular dynamics. Table 2 shows the main characteristics of the structure of the melts - distances between nearest neighbors (r), coordination numbers (z) and

Warren-Cowley parameters ($\alpha_{i-j}$) which were determined from the calculated total and partial radial distribution functions of atoms (RDF).

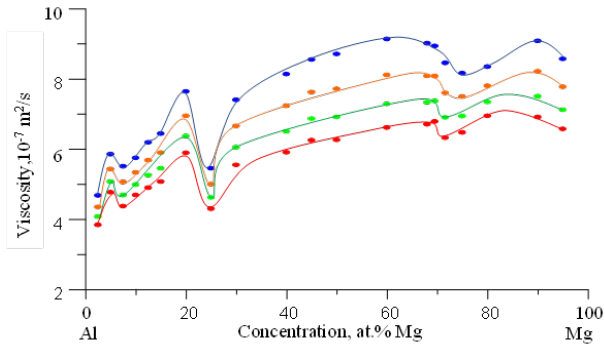

**Fig.2** Concentration dependences of the viscosity of Al-Mg melts at temperatures of 800 $^0$C (lower line), 750, 700 and 650 $^0$C (upper line).

The results are given in table. 2 were compared with the data obtained in [3] and are shown in fig. 3. Considering the advanced calculation methods we use, which allow us to carry out MD calculations with *ab initio* accuracy even for small concentrations of elements in alloys, the parameters $\alpha$ calculated by us are more accurate than the results obtained in 2017. The Cowley parameters obtained by us confirm the main regularities established in this article, but show that the chemical short-range order does not have such sharp changes as it was presented in [3].

**Table 2.** Distances between nearest neighbors ($r_{i-j}$), coordination numbers ($z_{i-j}$) and Warren-Cowley parameters ($\alpha_{i-j}$) in $Al_{80}Mg_{20}$ and $Al_{20}Mg_{80}$ melts at 600 and 800 $^0$C.

| | $Al_{80}Mg_{20}$ | | | | | | $Al_{20}Mg_{80}$ | | | | | |
|---|---|---|---|---|---|---|---|---|---|---|---|---|
| | 600 $^0$C | | | 800 $^0$C | | | 600 $^0$C | | | 800 $^0$C | | |
| | $r_{i-j}$ | $z_{i-j}$ | $\alpha_{i-j}$ | $r_{i-j}$ | $z_{i-j}$ | $\alpha_{i-j}$ | $r_{i-j}$ | $z_{i-j}$ | $\alpha_{i-j}$ | $r_{i-j}$ | $z_{i-j}$ | $\alpha_{i-j}$ |
| total | 2.83 | 12.4 | - | 2.81 | 12.0 | - | 3.07 | 12.5 | - | 3.03 | 12.3 | - |
| Al-all | 2.81 | 12.3 | - | 2.79 | 11.9 | - | 2.87 | 11.3 | - | 2.89 | 10.7 | - |
| Mg-all | 2.97 | 14.2 | - | 2.99 | 14.1 | - | 3.11 | 13.0 | - | 3.11 | 13.0 | - |
| Al-Al | 2.77 | 9.0 | 0.08 | 2.77 | 9.0 | 0.05 | 2.75 | 2.0 | 0.13 | 2.71 | 1.9 | 0.12 |
| Al-Mg | 2.97 | 2.8 | -0.16 | 2.95 | 2.8 | -0.18 | 2.93 | 9.5 | -0.04 | 2.93 | 9.2 | -0.08 |
| Mg-Al | 2.97 | 11.4 | 0 | 2.95 | 11.3 | 0 | 2.93 | 2.3 | 0.09 | 2.93 | 2.3 | 0.11 |
| Mg-Mg | 3.11 | 3.1 | -0.10 | 3.13 | 3.2 | -0.13 | 3.13 | 10.8 | -0.04 | 3.13 | 10.7 | -0.03 |

……………………………………………………………………………………..

The found concentration features of viscosity are in good agreement with the data on the concentration change in the chemical short-range order in Al-Mg melts (Fig. 3). Comparing the obtained concentration dependences of viscosity and Warren – Cowley parameters [3] (Fig. 2, 3), one can observe that the concentration dependence of viscosity reflects changes in the chemical short-range order in melts. A sharp change in the chemical interaction in the melt between Mg atoms is observed in the concentration range from 10 to 20 at.% Mg (Fig. 3), while the conglomeration of Mg atoms occurs, as a result of which the interaction between Al-Mg atoms weakens, which was pronounced at a minimum content mg. A similar feature with respect to Al atoms is also observed at 80 at.% Mg. Both of these features are manifested in the concentration dependences of viscosity.

For the geometric analysis of the structure of $Al_{80}Mg_{20}$ and $Al_{20}Mg_{80}$ melts, the method of rotational invariants was used. Presented in fig. 4 distributions were compared with the values of rotational invariants for the main types of close-packed structures, which have the following values: for fcc – W6 = -0.01316, q6 = 0.5745; for hcp – W6 = -0.01244, q6 = 0.4847; for the icosahedral phase (ico) – W6 = -0.1697, q6 = 0.6633. The conducted studies show that both liquid alloys have a close tendency to cluster formation and it does not undergo significant changes with a decrease in the melt temperature. The centers of formation of fcc and ico structures are both Mg and Al atoms. However, a more pronounced ico atomic order is formed around the Al atoms, even at its low content in the alloy.

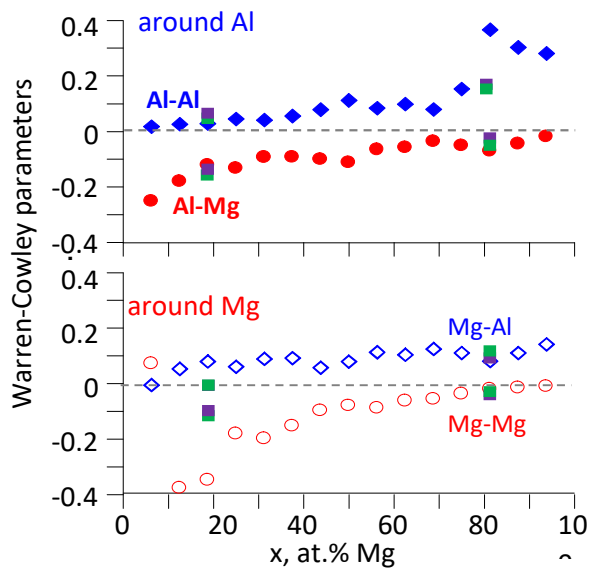

**Fig.3** Warren-Cowley parameters: data of this work (dark squares - 600 ⁰C, light squares - 800 0C) and work [3] (circles and rhombuses - 700 ⁰C).

In connection with the identified features of cluster formation, it is interesting to check how much such a structure of the melt affects the formation of crystalline structures in processes close to foundry and metallurgical ones. Therefore, we have studied the concentration dependence of undercooling and the processes of crystallization of Al-Mg alloys under DTA conditions. The temperatures of various stages of crystallization extracted from the DTA thermograms against the background of the equilibrium state diagram are shown in Fig. 5. It can be seen from the figure that the crystallization of alloys begins with a slight overcooling. Such a small amount of undercooling is associated with the high wettability of corundum ceramics by Al-Mg melts, the more Mg atoms in the melt, the higher the wettability. However, even at the observed low values of ΔT, it can be seen that at Mg concentrations of more than 69 at.%, supercooling increases, and for $Al_{90}Mg_{10}$ and $Al_5Mg_{95}$ alloys, a nonequilibrium character of crystallization is observed, at which the eutectic type of crystallization is preserved. As well as on the DTA heating curves, the cooling curves do not show phase transformations associated with the decomposition of intermediate phases (ζ-$Al_{52}Mg_{48}$ [6] and ε-$Al_{30}Mg_{23}$ [7]), on the cooling thermograms of alloys with Mg content from 45 to 60 at. % one exothermic effect is observed. An analysis of the microstructure shows that for the majority of the investigated alloys, despite the small values of supercooling, under which crystallization begins, the formation of nonequilibrium structures is typical. Alloys $Al_{90}Mg_{10}$ and $Al_{10}Mg_{90}$ solidify with the formation of eutectics, which indicates the preservation of the liquid phase upon cooling to eutectic temperatures (450 and 438 ⁰C).

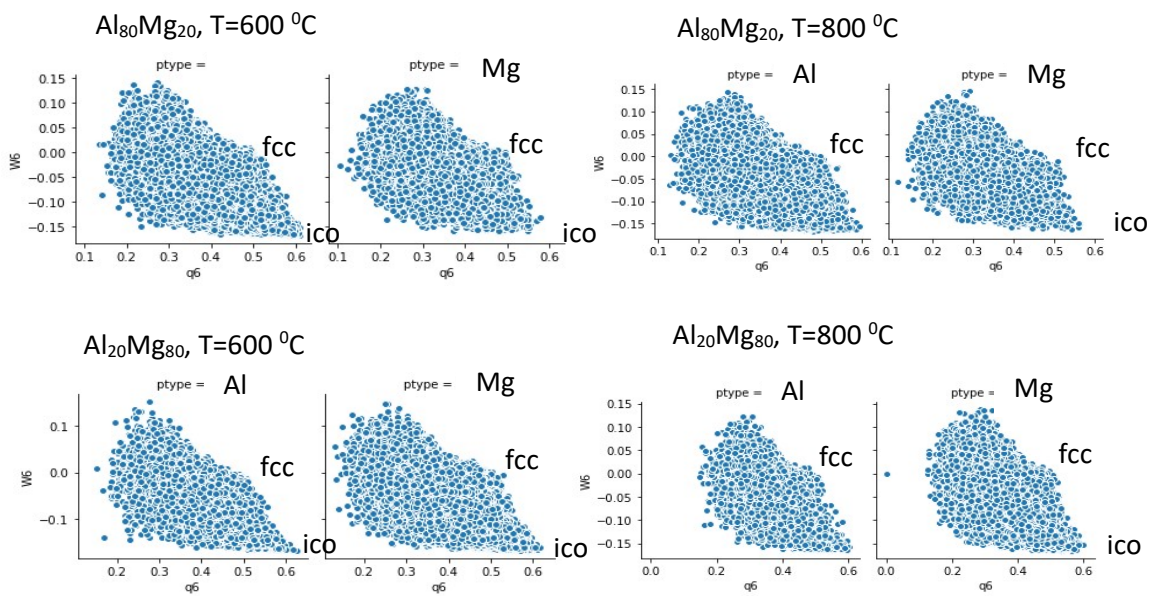

**Fig.4** Correlators (joint probability density function) of the bond-orientational order parameters BOOP, W6 and q6, for each atom type in the Al-Mg melts.

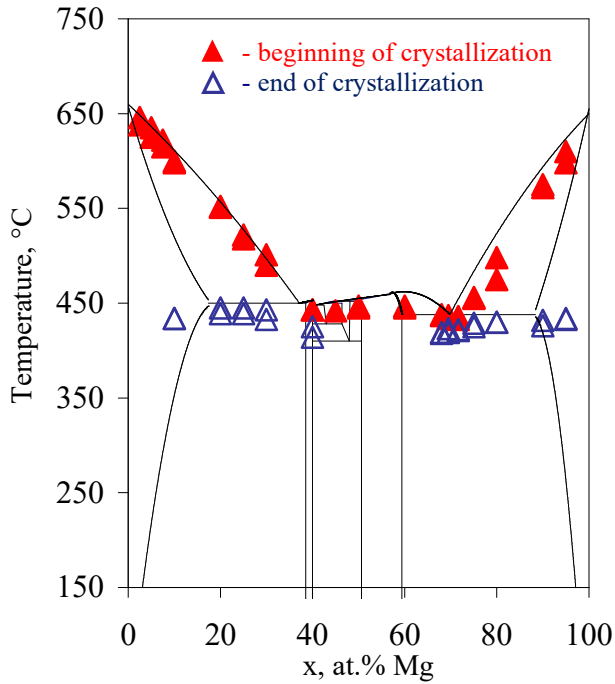

**Fig.5** Equilibrium state diagram of the Al-Mg system [22] with temperatures of exothermic effects according to DTA data upon cooling (v = 100°C/min) of Al-Mg alloys from 800°C.

During equilibrium crystallization in $Al_{90}Mg_{10}$ and $Al_{10}Mg_{90}$ alloys, only solid solutions based on Al and Mg should form from the liquid phase, and crystallization should end at temperatures above 500 $^0C$, after which, upon further cooling, as a result of a change in the solubility of alloying components, secondary intermetallic compounds should be released from solid solutions during solid phase reactions. The nonequilibrium crystallization observed by us testifies to the difficulty of the diffusion redistribution of Mg in the Al matrix for the $Al_{90}Mg_{10}$ alloy and Al in the Mg matrix for the $Al_{10}Mg_{90}$ alloy, as a result of which the growth of the solid solution is limited at a certain stage. Qualitative analysis of the ratio between the areas of eutectic components and solid solutions indicates that the formation of a Mg-based solid solution stops much earlier than that of an Al-based solid solution.

The DTA thermograms were used to calculate the values of the supercooling required to initialize the crystallization process, and plot its concentration dependence (Fig. 6). From fig. Figure 6 shows that when the studied melts are cooled at rates of 20, 50 and 100 $^0C$/min from different temperatures, a large spread in the overcooling value is observed. Increasing the cooling rate for most alloys leads to an increase in subcooling, however, these changes are not as significant as when varying the temperature.

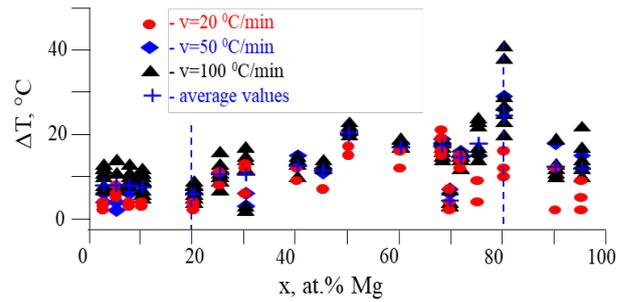

**Fig.6** Concentration dependences of undercooling of Al-Mg melts.

Despite the scatter in the values of ΔT, the concentration dependence of undercooling shows concentrations at which a fundamental change in the nature of the concentration behavior of ΔT occurs, which is not associated with a change in the type of crystals formed (blue dotted lines in Fig. 6). In the concentration range from 2.5 to 40 at.% Mg at undercoolings up to 20 $^0C$, crystallization begins with the formation of an Al-based solid solution, and up to 20 at.% Mg, the effect of concentration on the magnitude of supercooling is not observed (Fig. 6). When the magnesium content is from 20 to 40 at.%, the supercooling value increases (Fig. 6), which indicates a decrease in the crystallization ability of the Al-based solid solution.

The formation of $Al_3Mg_2$ intermetallide (40-45 at.% Mg) also proceeds at low undercoolings ~10 $_0C$. The $Al_{12}Mg_{17}$ intermetallic compound is formed first from the melt in the concentration range from 50 to 70 at.% Mg; its formation requires greater supercooling than for the $Al_3Mg_2$ intermetallic compound; therefore, at 50 at.% Mg, a jump is observed in the concentration dependence of ΔT (Fig. 6). A further increase in the Mg content in the melt increases the crystallization capacity of $Al_{12}Mg_{17}$. The minimum undercooling values for the Al-Mg system (~5 $^0C$) are observed for the $Al_{30}Mg_{70}$ alloy - this is a eutectic alloy, but under the selected cooling conditions, its crystallization begins with the growth of primary Mg dendrites. Thus, the Mg-based solid solution is the first from the melt to solidify in the concentration range from 70 at.% Mg and more. In the concentration range from 70 to 80 at.% Mg, undercooling increases, and then, with an increase in the Mg concentration to 95 at.%, it decreases. From the analysis of the concentration dependence of undercooling, two concentrations, 20 and 80 at.% Mg, can be distinguished, at which there is a significant change in undercooling, but no change in the type of crystallization is observed.

The found concentration features of undercooling are in good agreement with the data on the concentration change in viscosity and chemical short-range order in Al-Mg melts [3] (Fig. 2, 3).

The strong effective attraction between Mg atoms and the resulting local segregation can explain the nonequilibrium crystallization of the melt observed by us, in which the liquid phase is retained until eutectic crystallization. Features of phase formation in Al-Mg alloys at Mg concentrations of more than 20 at.% are mainly determined by the interaction between Al atoms. Al atoms, as they are replaced by Mg, tend to maintain interaction with each other, as a result of which they shift to more distant distances than the first coordination sphere [3], which leads to a decrease in their number both around Al and Mg atoms. This interaction is maximally manifested at an equiatomic ratio of Al and Mg and in the region of 80 at.% Mg. During crystallization, this interaction prevents the formation of disordered phases (solid solutions based on Al and Mg) and contributes to an increase in undercooling, under which their crystallization is observed.

Thus, the conducted studies have shown that the viscosity of Al-Mg melts in the temperature range from the melting point to 800 $^0$C decreases exponentially with increasing temperature. The concentration dependence of the viscosity of Al-Mg melts from 2.5 to 95 at.% Mg is not monotonic and has pronounced features in the region of 20 and 80 at.% Mg. Studies of the processes of crystallization of Al-Mg alloys have shown that the crystallization of most alloys begins at small supercoolings (from 5 to 20 $^0$C) and depends on the cooling conditions and concentration, the maximum supercooling is achieved for the $Mg_{80}Al_{20}$ alloy when cooled from 700 $_0$C and can reach 40 $_0$C. The concentration dependence of undercooling has features in the regions of 20, 35, 50, 70, and 80 at.% Mg. Features at 35, 50 and 70 at.% Mg are associated with a change in the crystal structure formed from the phase melt. The breaks in the concentration dependence of undercooling at 20 and 80 at.% Mg correspond to the concentrations at which the maximum chemical interactions in melts are observed (at 20 at.% Mg - effective attraction of Mg-Mg, at 80 at.% Mg - effective repulsion of Al-Al ).

## Acknowledgements

The investigation has been financed by the grant of the Russian Science Foundation No. 22-22-00912, https://rscf.ru/project/22-22-00912/.

## References

[1] Lin Wang, Shanshan Li, Lin Bo, Di Wu, Degang Zhao Liquid-liquid phase separation and solidification behavior of Al-Bi-Sn monotectic . alloy. Journal of Molecular Liquids 254 (2018) 333–339 (https://doi.org/10.1016/j.molliq.2018.01.118)

[2] O.S. Roik, O.V. Samsonnikov, V.P. Kazimirov, V.E. Sokolskii, S.M. Galushko Medium-range order in Al-based liquid binary alloys Journal of Molecular Liquids 151 (2010) 42–49

[3] Jin Wang , Xinxin Li, Shaopeng Pan, Jingyu Qin, Mg fragments and Al bonded networks in liquid Mg–Al alloys. Computational Materials Science 129 (2017) 115–122. (http://dx.doi.org/10.1016/j.commatsci.2016.12.006)

[4] Young-Ki Yang, Todd Allen. Direct visualization of β phase causing intergranular forms of corrosion in Al–Mg alloys // Materials Characterization, 2013, V. 80, pp. 76 – 85.

[5] Ki Yang, Todd Allen. Direct visualization of β phase causing intergranular forms of corrosion in Al–Mg alloys // Materials Characterization, 2013, V. 80, pp. 76 – 85.

[6] J.S. Kasper, R.M. Waterstrat, Ordering of atoms in the r phase, Acta Cryst. 9 (3) (1956) 289–295.

[7] S. Samson, E.K. Gordon, The crystal structure of e-Mg23Al30, Acta Crystallogr. B 24 (8) (1968) 1004–1013.

[8] B.L. Mordike, T. Ebert, Magnesium: properties–applications–potential, Mater. Sci. Eng. A 302 (2001) 37–45.

[9] Y. Zhong, M. Yang, Z.K. Liu, Contribution of first-principles energetics to Al–Mg thermodynamic modeling, Calphad 29 (4) (2005) 303–311 S.

[10] Vrtnik, S. Jazbec, M. Jagodic̆, A. Korelec, L. Hosnar, Z. Jaglic̆ić′, P. Jeglic̆, M. Feuerbacher, U. Mizutani, J. Dolinšek, Stabilization mechanism of c-Mg17Al12 and b-Mg2Al3 complex metallic alloys, J. Phys.: Condens. Matter 25 (42) (2013) 425703

[11] D. Shin, C. Wolverton, The effect of native point defect thermodynamics on offstoichiometry in b-Mg17Al12, Acta Mater. 60 (13) (2012) 5135–5142.

[12] Debela T.T., Abbas H.G. Role of nanosize icosahedral quasicrystal of Mg-Al and Mg-Ca alloys in avoiding crystallization of liquid Mg: Ab initio molecular dynamics study // Journal of Non-Crystalline Solids,  2018, V. 499 pp.173-182.

[13] Kamaeva L.V., Ryltsev R.E., Lad'yanov V.I., Chtchelkatchev N.M. Viscosity, undercoolability and short-range order in quasicrystal-forming Al-Cu-Fe melts // Journal of Molecular Liquids, 2020, V. 299, pp. 112207(9). DOI: 10.1016/j.molliq.2019.112207

[14] Kamaeva L.V., Sterkhova I.V., Lad'yanov V.I., Ryltsev R.E., Chtchelkatchev N.M. Phase selection and microstructure of slowly solidified Al-Cu-Fe alloys // Journal of Crystal Growth, 2020, V. 531, pp. 125318. DOI: 10.1016/j.jcrysgro.2019.125318

[15] Kamaeva L.V., Ryltsev R.E., Suslov A.A., Chtchelkatchev N.M. Effect of copper concentration on the structure and properties of Al–Cu–Fe and Al–Cu–Ni melts // Journal of Physics: Condensed Matter, 2020, V. 32, pp. 224003(9). DOI: 10.1088/1361-648X/ab73a6

[16] Pepelyaeva V.D., Kamaeva L.V., Ryltsev R.Ye., Shchelkachev N.M. Influence of chemical interaction on the features of the solidification of the melts Al-Cu-Ni with 10 at.% Ni at small cooling rates // Khimicheskaya fizika i mesoskopiya [Chemical physics and mesoscopy], 2021, V. 23, No. 1, pp. 90-100. DOI: 10.15350/17270529.2021.1.9

[17] A.L. Bel'tyukov, V.I. Lad'yanov, An automated setup for determining the kinematic viscosity of metal melts, Instrum. Exp. Tech. 51 (2008) 304–310, https://doi.org/10.1134/S0020441208020279.

[18] E.G. Shvidkovskiy, Certain problems related to the viscosity of fused metals. NASA technical translation F-88,

1962, Translation of Nekotoryye Voprosy Vyazkosti Rasplavlennykh Metallow, State Publishing House for Technical and Theoretical Literature, Moscow, 1955 , (in Russian).

[19]    I.V. Sterkhova, L.V. Kamaeva, V.I. Lad'yanov, On the viscosity of Fe79Cr21 melt of congruent composition, Bull. Russ. Acad. Sci. Phys. 72 (10) (2008) 1365–1367, https://doi.org/10.3103/S106287380810016X.

[20]    Sterkhova I.V., Kamaeva L.V. The influence of Si concentration on undercooling of liquid Fe // J. Non-Cryst. Solids, 2014, V. 401, pp. 250–253.

[21]    G. Kresse, J. Furthmuller, Efficiency of ab-initio total energy calculations for metals and semiconductors using a plane-wave basis set, Comput. Mater. Sci. 6 (1) (1996) 15–50, https://doi.org/10.1016/0927-0256(96)00008-0.

[22]    J.L. Murray, The Al−Mg (Aluminum−Magnesium) system. Bulletin of Alloy Phase Diagrams, 1982, V. 3(1) pp. 60 - 74.